\documentclass[twocolumn,showpacs,aps,prd,nobibnotes,nofootinbib,floatfix,superscriptaddress,longbibliography]{revtex4-1}

\usepackage{amsmath,amssymb,bm,xfrac}
\usepackage{graphicx}
\usepackage[usenames]{xcolor}
\usepackage[normalem]{ulem}
\usepackage{xspace}
\usepackage{dsfont}
\usepackage{soul} % Strike-throughs and underline support
\usepackage{dsfont}
\usepackage{multirow}
\usepackage{float}
\usepackage{booktabs}
\usepackage{cancel}
\DeclareUnicodeCharacter{2212}{-}
\usepackage[caption=false]{subfig}
\usepackage[colorlinks=true,allcolors=blue]{hyperref}

\usepackage[caption=false]{subfig}

\usepackage{hyperref}%
\usepackage{appendix}
\usepackage{xcolor}
\newcommand\araa{{ARA\&A}}%  % Annual Review of Astron and Astrophys 
\renewcommand\apj{{ApJ}}%    % Astrophysical Journal 
\newcommand\apjl{{\@eapj@ApJLetters}}%     % Astrophysical Journal, Letters 
\newcommand\apjs{{ApJS}}%    % Astrophysical Journal, Supplement 
\newcommand\aap{{A\&A}}%     % Astronomy and Astrophysics 
\newcommand\mnras{{MNRAS}}%   % Monthly Notices of the RAS 
\renewcommand\prc{{Phys.~Rev.~C}}% % Physical Review C 
\renewcommand\prd{{Phys.~Rev.~D}}% % Physical Review D 
\renewcommand\prl{{Phys.~Rev.~Lett.}}% % Physical Review Letters 
\newcommand\ssr{{Space~Sci.~Rev.}}% % Space Science Reviews 
\newcommand\nphysa{{Nucl.~Phys.~A}}%     % Nuclear Physics A 
\newcommand\physrep{{Phys.~Rep.}}%       % Physics Reports 
\newcommand\physscr{{Phys.~Scr}}%        % Physica Scripta 
\newcommand\pasa{{PASA}}%  % Publications of the Astron. Soc. of Australia
\newcommand{\orcid}[1]{\href{https://orcid.org/#1}{\includegraphics[height=\fontcharht\font`\B]{Figures/ORCIDiD_icon16x16.png}}}

\newcommand{\JHU}{William H. Miller III Department of Physics \& Astronomy, Johns Hopkins University, 3400 N Charles St, Baltimore, MD 21218, USA}
\newcommand{\ARCO}{Astrophysics Research Center of the Open University (ARCO), \\Department of Natural Sciences, Ra’anana 4353701, Israel}
\newcommand{\Salamanca}{Department of Fundamental Physics and IUFFyM, University of Salamanca,\\ Plaza de la Merced S/N E-37008, Salamanca, Spain}
\newcommand{\Oxford}{Department of  Physics, University of Oxford, Keble Road OX1 3RH, Oxford, UK}
\newcommand{\CNRS}{Institut d’Astrophysique, UMR 7095 CNRS, \\Sorbonne Universit\'e, 98bis Blvd Arago, 75014 Paris, France}

\begin{document}

\title{Neutrino diagnostics of hadron-quark phase transition in Neutron Stars}

%\author[0000-0002-0632-8897]{Yossef Zenati}
\author{Yossef Zenati}%~\orcid{0000-0002-0632-8897}}
 \email{yzenati1@jhu.edu}
\affiliation{\JHU}
\affiliation{\ARCO}
 
\author{Conrado Albertus Torres}
 \email{albertus@usal.es}
\affiliation{\Salamanca}

\author{Joseph Silk}
 \email{silk@iap.fr}
\affiliation{\Oxford}
\affiliation{\CNRS}
\affiliation{\JHU}

\author{ M. \'Angeles P\'erez-Garc\'ia}
 \email{mperezga@usal.es}
 \affiliation{\Salamanca}

\date{\today}

\begin{abstract}
We investigate neutrino signatures of a hadron–quark phase transition (HQPT) in neutron stars (NS)  leading to quark star (QS)  formation. %Deconfinement is triggered once the central density exceeds a critical threshold above $\sim 3\,n_0$ being $n_0$, saturation density. 
We use representative hadronic and quark equations of state i.e.  DD2 and MIT bag model along with a phenomenological neutrino-emission model including the dominant leptonic and hadronic processes.
%We account for neutrinos uage scheme with General Relativity (GR) redshift. 
Rather than aiming at a fully consistent hydrodynamical simulation, our goal is to identify generic temporal diagnostics that may arise when deconfinement occurs during the evolution of a compact star. We identify characteristic diagnostic features that may emerge in the neutrino light curve on $\simeq 10–50$ ms timescales. These include an enhanced peak-to-plateau ratio, a delay tracing the central-density evolution, and a transient spectral hardening. %(i) an enhanced peak-to-plateau ratio, (ii) a time lag between the collapse rise and the deconfinement signal that traces the central-density evolution, and (iii) a transient hardening of the neutrino spectrum driven by quark-matter phase space. 
After standard MSW flavor conversion, these temporal and spectral signatures remain potentially detectable for Galactic events under optimistic assumptions with detectors such as IceCube and Hyper-Kamiokande. Our results suggest possible temporal and spectral diagnostics of quark deconfinement in future Galactic neutrino bursts.

%After MSW flavor conversion, these signatures remain detectable with current experiments; for a Galactic event ($d\sim10$ kpc), IceCube and Hyper-K should resolve the HQPT feature and distinguish it from both no-transition NS collapse and canonical core-collapse supernova(CCSN) templates. 
\end{abstract}


\maketitle

%%%%%%%%%%%%%%%%%%%%%%%%%%%%%%%
%%% Section 1: Introduction %%%
%%%%%%%%%%%%%%%%%%%%%%%%%%%%%%%
{\bf Introduction}.--%\label{sec:introduction}
A long-standing prediction of Quantum Chromodynamics (QCD) for cold or warm ultradense matter is that, above a few times nuclear saturation density, $n_0$, hadrons may deconfine into their quark constituents and form a new phase of strongly interacting matter \citep{Alford+08, Lattimer_Prakash04,Baym+18,YangY+25}. The cores of (proto-) neutron stars (NSs) are considered potential astrophysical sites for this transition, provided that the thermodynamical conditions are favorable. Further, nucleation of quark bubbles due to exotic external Beyond Standard Model (BSM) agents, such as self-annihilating dark matter accumulating in the NS cores has been quoted as yet another mechanism triggering the transition \cite{Perez_Garcia_2010,Herrero_2019}.  Depending on the equation of state (EOS) and the microphysical realization of the transition, the macroscopic outcome could range from a hybrid star with a quark core to a complete conversion into a self-bound quark star (QS) \citep{Bombaci+16, AnnalaE+20}. One central fact is related to the behavior of the speed of sound $c_s$ in strongly interacting matter, aiding the search to identify a QS \citep{McLerran_Reddy2019,AnnalaE+20,MarczenkoM24}.

Identifying observable neutrino signatures associated with a hadron–quark phase transition (HQPT) in compact stars remains a major challenge. Various studies have suggested that such transitions could leave imprints in astrophysical observables \citep{BausweinA+19, McLerran_Reddy2019,BlackerS+20,TakatsyJ+23}, motivating the search for diagnostic features in neutrino signals.

%%%%%%%%%%%%%%%%%%%%%%%%%%%%%%%%%%%%%%%%%%
\begin{figure*}[t]
    \centering
    \includegraphics[width=0.99\textwidth]{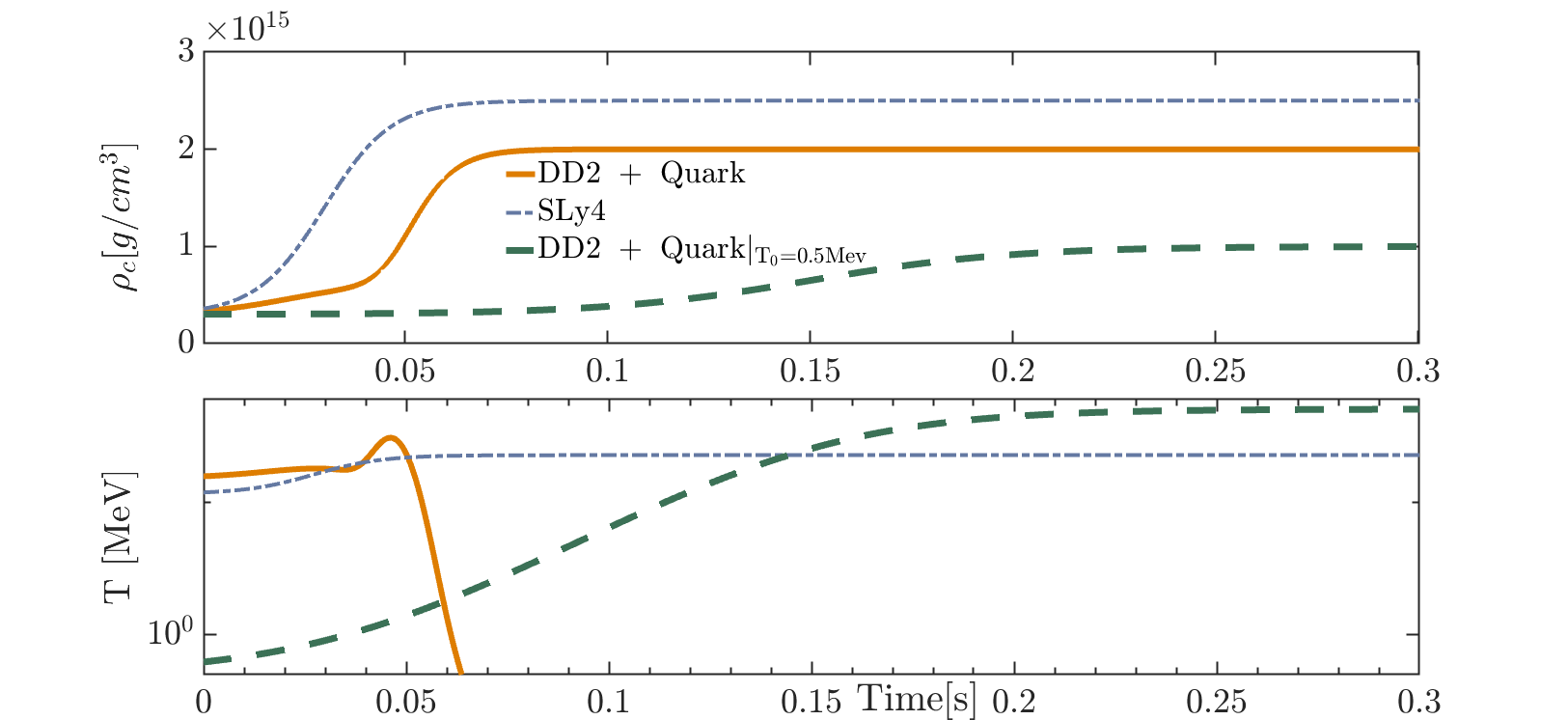}
    \caption{Central density (upper panel) and temperature (lower panel) time evolution in a NS collapsing to QS under the DD2 EOS set at initial $\rm k_BT_{0} = 0.5$ MeV (green dashed line) and warm $\rm k_BT_{0} = 10$ MeV (orange curve). As a comparison we plot the warm initial case with alternative  SLy4 \citep{sly41998NuPhA.635..231C} hadron EOS (blue dash line). For the latter no transition to QS arises culminating in prompt black hole formation instead. Quark matter is  described always by MIT bag model. }
    \label{fig:DD2_SLy4}
\end{figure*}
%%%%%%%%%%%

A transient HQPT can imprint characteristic time-domain and spectral signatures in the neutrino signal, a prompt rise or secondary “burst” changes in flavor hierarchy, and a hardening of the mean energies—features that are, in principle, accessible to current and next-generation detectors such as IceCube\citep{Aartsen:2016nxy}, Super-K/Hyper-K \citep{HyperK2018}, KM3NeT \cite{Adrian-Martinez:2016fdl}, and JUNO \cite{AdamT_JUNO2015}
for Galactic events \citep{ScholbergK12}.

However, neutrino signatures of NS to QS conversion outside the Supernova context remain largely unexplored. Some other lie in the contexts of proto–neutron star cooling, rotating NS  instabilities from  angular momentum loss \cite{dimmel2009MNRAS.396.2269D},  core collapse supernovae (CCSN)  \cite{Khosravi_Largani_2024} where the onset density of a possible phase transition is constrained from a future neutrino observation, and merger remnants \citep{Nakazato+08, FischerT+18, BausweinA+19}. A dedicated microphysics-based analysis focused on NS to QS conversion scenarios remains relatively unexplored \citep{AnnalaE+20}. 

In binary NS (BNS) coalescence a phase transition to deconfined quark matter is expected to influence the postmerger gravitational wave (GW) emission from a merger remnant and constrain the regime of viable NS EOSs \citep{Most+18,AnnalaE+18,AnnalaE+22}. Recent works such as \citep{BausweinA+19,MostE+20, Combi_Siegel23,PrakashA+24} indicate that a strong first-order HQPT can imprint a distinctive shift in the post-merger GW. Transitions of higher order or even cross-over have also been discussed recently, see \cite{Tak_tsy_2023}. The dominant GW mode with kHz peak $f_2$ becomes an outlier relative to the otherwise tight empirical $f_2$ -$\tilde{\Lambda}$ relation, based on tidal polarizability $\tilde{\Lambda}$, inferred from the merger, providing a clean joint inspiral or post-merger diagnostic of deconfinement \citep{BausweinA+19}. Using a finite-temperature chiral mean-field EOS in full General Relativity (GR)  simulations, \citet{MostE+20} showed that deconfinement chiefly affects the {post-merger} stage—inducing GW dephasing, shifting $f_2$, and, for stronger transitions, hastening the collapse and truncating the kHz signal. Finally, end-to-end numerical Relativity (NR) plus Bayesian studies \citep{PrakashA+24} find that next-generation detectors such as the Einstein Telescope (ET) or Cosmic Explorer (CE) will be able to detect HQPT-induced $f_2$ deviations at post-merger SNR$\gtrsim 10$ for sufficiently strong transitions, whereas weaker or mixed-phase scenarios may remain degenerate due to hadronic uncertainties. Recently, one-dimensional numerical simulations using general relativistic neutrino radiation demonstrate that a phase transition occurs during the accretion-induced collapse of a white dwarf (WD), which results in a high neutrino burst\citep{ChanJ+24}.

In this work, we investigate whether a neutrino luminosity curve, by itself, can serve as a  potential diagnostic of QS formation during a HQPT. For this we adopt a spherically symmetric representative collapse configuration in GR \citep{OConnor+10_1dHydroGR} with an effective collapse evolution calibrated to representative GR collapse scenarios in which a thermalized NS  undergoes deconfinement once a critical density is exceeded, roughly $\sim3n_0$ (with $n_0 \simeq 2.2\times10^{14}\,\mathrm{g\,cm^{-3}}$). For the hadronic phase we employ the well-tested DD2 EOS \citep{Typel+10_EOS, HempelM+10_EOS, FischerT+11,FortinM+18_DD2} at fixed electron fraction $Y_e=0.2$ and warm core temperature $k_BT=10\,\mathrm{MeV}$. According to the current models of evolution these temperatures are typically achieved at 1-2 seconds. For comparison we also consider a colder situation around $k_BT=0.5\,\mathrm{MeV}$ typically obtained after 2-3 minutes  following the SN collapse at $Y_e\lesssim 0.1$.
We assume triggering conditions are fulfilled, either in the standard or exotic scenarios described, and analyze a canonical configuration with gravitational mass $\rm M_{NS}=1.4\,M_\odot$. Within this background we compute neutrino luminosities from the NS collapse into QS using composition and temperature-dependent neutrino emissivities responsible for the neutrino energy exchange. We track the early neutrino trapping phase through an approximate opacity-based treatment consistent with a 1D-GR leakage/transport scheme capturing the essential physics \citep{OConnor+10_1dHydroGR,OConnorOtt_11,Zenati_Albertus_Angeles_Silk23} although we keep in mind that the back reactions will not be captured as in state-of-the-art simulations \cite{Khosravi_Largani_2024}. Our aim is not to provide a detailed simulation of the conversion dynamics, but to isolate characteristic features that may arise in the neutrino signal when deconfinement occurs.

In this spirit our microphysics input includes the dominant channels relevant before and after quark deconfinement with their corresponding opacity and phase-space factors \citep{Iwamoto82,YakovlevD+01,PotekhinAY+15,FiorellaBurgio+18,Schatz+22}, see Table \ref{tab:table1} in appendix\ref{app:microphysics_table}. 

In more detail, we include the modified Urca process which conserves momentum in regimes where the direct Urca process is forbidden. 
Charged-current absorption and neutral-current scatterings on the relevant constituents (nucleons or quarks) are included to determine optical depths and the transition between diffusion and free-streaming–dominated regimes. This setup enables a transparent mapping from microphysical assumptions to observable features in the time-dependent neutrino luminosity and mean energies for each flavor, see Table \ref{tab:table1}  in appendix \ref{app:microphysics_table} for actual reaction rates, emissivity, opacities \cite{FrimanB_MaxwellV79,MeyerS94,RaffeltG1996,HannestadS_RaffeltG98,ThompsonT+00,KohriK_MineshigeS02,KohriK+05} and alternatively for quarks \cite{IwamotoN82_quark,FischerT+11, SchmittA_ShterninP18,ShahrbafM+20,Rosero-Gil_LugonesG21}. In our standard scenario, we do not include any explicit exotic energy carriers, although sterile neutrinos and other BSM species are, in principle, not excluded and have been proven to yield distinctive features \cite{Albertus_2015,Rembiasz_2018}.

{\bf Neutron Star Collapse to a Quark Star}.--
\label{sec:methods}
%In this section, we provide  the general relativistic hydrodynamic framework relevant to the collapse of a NS transitioning to a quark star phase.
We consider an initial NS matter configuration where only nucleons (protons and neutrons) along with a component of electrons,  muons and neutrinos form an electrically neutral system. We select the DD2 EOS \citep{Typel+10_EOS} for the range of baryonic number densities $n_B\in[10^{-9}, 1.5]$ $\rm fm^{-3}$ at reference cases $k_BT\sim [0.5,10]$ MeV. %Saturation density is fixed at $n_0=0.149\,\rm fm^{-3}$ (in mass density $\rho_0\simeq 2.5\times 10^{14}\rm g/cm^3$).
As known, due to increasing complexity, description of the different NS layers is not based on the same degrees of freedom but it effectively produces phenomenological configurations within the current nuclear and astrophysical constraints. The NS maximum mass in this setting is $M\sim \rm 2.42M_\odot$ with a radius of $R\sim 11.95$ km and for $M=\rm 1.4M_\odot$,  $R_{1.4}\sim 13.19$ km \citep{EssickR+20, GambaR+20}. The tidal deformability for the simulated BNS merger event producing GW170817 at a mass ratio $q \sim 0.8$ is $\Lambda\sim 783.05$ \citep{Most+18,DeS+18,HanSophia+19}. Our setup assumes spherical symmetry and incorporates essential GR corrections relevant for strong-field gravity, neutrino redshift, and quark matter nucleation. %We briefly describe the GR simulation and the metric we used in section \ref{sec:simGR}.

Inside the NS and above a critical transition mass density $\rho_t$, the system evolves to a deconfined quark state, that we model by the MIT bag model
\citep{Chodos+74_MITbag,JohnsonK_MITbag_78,DexheimerV+13,LopesL+21,LopesL+25}. Within this framework the pressure of quark matter is obtained by adding the partial pressure of each quark species (u,d,s) and $B$ is the bag constant 
encoding the QCD vacuum energy contribution. For details, we refer to appendix\ref{app:mitbag}. We take $B^{1/4} \simeq \rm 130$ MeV %regarding soft quark matter EOS with low temperature 
\citep{BordbarH+06,BombaciI+11,OertelM+17,LopesL+25} as a representative value. %In order to approximate the complex partially deconfined NS interior

In our setup, the transition region obtained in the simulation is represented through a smooth interpolation \citep{AyriyanA+18,ShahrbafM+20} governed by a $\chi$ function

\begin{equation}
\chi(r,t) = \frac{1}{2} \left[ 1 + \tanh\left(\frac{\rho(r,t) - \rho_t}{\Delta \rho}\right)\right].
\end{equation}

$\Delta \rho$, the width parameter, characterizes the density interval across which the deconfinement front develops. Physically, the finite transition width reflects unresolved microphysics associated with the hadron–quark interface, including finite-size and surface-tension effects. Thus the pressure is interpolated as

\begin{equation}
P(r,t) = (1 - \chi) P_{\mathrm{h}} + \chi P_{\mathrm{q}},
\end{equation}

where $\chi(r,t)$ describes the local fraction of quark matter. $P_{\mathrm{h}}, P_{\mathrm{q}}$  are the hadron and quark pressure contributions, respectively.

The present effective interpolation is intended to capture finite-width transition effects beyond the idealized sharp-interface limit. It allows us to assess the sensitivity of the predicted neutrino 
signal to a non--abrupt deconfinement transition. A crossover transition, lacking phase coexistence, remains an equally plausible alternative.

As the quark phase develops in the stellar interior, the energy release associated with the 
deconfinement transition can be approximated by $L_q(t) = \frac{dV_q(t)}{dt}\,(\epsilon_q - \epsilon_h)$, where $V_q(t)$ denotes the stellar volume converted into quark matter and $\epsilon_{q,h}$ are the local 
energy densities of the quark and hadronic phases, and $\Delta \epsilon = \epsilon_q - \epsilon_h$ its difference, respectively. 
This expression retains only the bulk contribution from the energy--density difference, neglecting 
finite--size and thermal corrections, and therefore provides a leading--order estimate of the local 
energy release that can feed a transient neutrino burst.

To relate the local emission to the signal observed at infinity, we adopt a static, spherically symmetric background  so that the observed luminosity reads $L_\infty = \alpha^2(r)\,L(r)$ with $\alpha(r) = \sqrt{1-\frac{2GM(r)}{rc^2}}$ the lapse function of the Schwarzschild geometry. 
%The quadratic dependence reflects the combined effect of gravitational energy redshift and time 
%dilation acting on the neutrino flux.

In more detail, we model the time--dependent neutrino luminosity associated with the conversion and for this we consider representative initial stellar configurations 
with temperatures  $T = [0.5,10]\,\mathrm{MeV}$ ($k_B=1$) and electron fraction $Y_e \lesssim[ 0.2,0.1]$. The neutrino emission is computed from the relevant microphysical processes, shown in Table \ref{tab:table1} in appendix\ref{app:microphysics_table}. 
%includes both conventional nucleon--mediated channels and an additional contribution associated 
%with the onset of deconfinement.

The total energy release associated with the conversion can then be written as $
\frac{dE}{dt}\sim L_q(t)$. 

The conversion dynamics in our leakage-based evolution exhibit an approximately sigmoidal growth of the quark-matter volume centered at $t_q = 45\,\mathrm{ms}$ with 
characteristic width $\tau_q = 10\,\mathrm{ms}$,

\begin{equation}
Sig(t) = \frac{1}{1+\exp\!\big[-(t-t_q)/\tau_q\big]}.
\end{equation}

The converted volume is then written as $V_q(t) = V_f\,Sig(t)$, where $V_f$ is the final 
quark--matter volume. Its time derivative reads

\begin{equation}
\frac{dV_q}{dt} = \frac{V_f}{\tau_q}\,Sig(t)\,[1-Sig(t)],
\end{equation}

leading to an energy release rate
$\frac{dE}{dt} = \Delta\epsilon \, \frac{V_f}{\tau_q}\,Sig(t)\,[1-Sig(t)]$. The maximum occurs at $t=t_q$, where $Sig=1/2$, yielding $\frac{dE}{dt}|_{\rm peak} = \frac{\Delta\epsilon\,V_f}{4\tau_q}$.

This behavior reflects the finite-timescale saturation of the conversion front observed in the simulation. At the coexistence point, where the two phases 
share the same pressure and chemical potentials, $\Delta\epsilon$ approximately corresponds to the latent 
heat per unit volume. Away from coexistence the interpretation is less direct, and the above 
expression should be regarded as an effective estimate of the bulk energy release.

We stress that the quantity 
\(
\Delta\epsilon\,\frac{dV_q}{dt}
\)
should not be interpreted as the neutrino luminosity produced by the phase transition. 
In realistic neutrino–radiation hydrodynamics simulations the released latent heat 
is redistributed among several channels, including changes in the internal energy, 
hydrodynamic work, and modifications of the local temperature and composition. 
Neutrino emission then arises indirectly through weak processes whose rates depend 
on these thermodynamic conditions and on neutrino transport. A fully consistent 
treatment of the conversion dynamics would require neutrino–radiation hydrodynamical 
simulations with detailed weak interaction rates \cite{hempel2009PhRvD..80l5014H}, which lies beyond the scope of 
the present work.

Each channel's emissivity is calculated in CGS units (erg\,cm\(^{-3}\)\,s\(^{-1}\)) and integrated over the NS volume. Neutrino trapping is included via an exponential suppression factor $\sim\exp(-\tau)$, where $\tau \propto \rho^{1.5}$ represents a simplified optical depth model growing with central density\citep{RaffeltG1996,FischerT+11,RobertsF+12,Roberts_Reddy17}. The behaviour of thermodynamical quantities are driving the ultimate neutrino dynamics. In Fig. \eqref{fig:DD2_SLy4} we compare central density and temperature under two different assumed EOS, i.e. the quark deconfinement through  DD2+MIT at initial temperatures $T_0=0.5,10$ MeV and an alternative fully nucleonic EOS, SLy4 \citep{sly41998NuPhA.635..231C}. The DD2+MIT model exhibits a sharp increase in central temperature and pressure around ($t \sim 0.045\,\mathrm{s}$), associated with the quark deconfinement phase transition, while the SLy4 model shows a smooth and rapid rise in all quantities, culminating in prompt black hole formation without a quark transition.

{\bf Simulation and neutrino signal in NS collapse to QS.--} In the GR formulation to obtain the neutrino signal from our simulations we adopt the Misner-Sharp metric for spherically symmetric with usual $t,r,\theta$ coordinates

\begin{equation}
ds^2 = -\alpha^2(t, r), dt^2 + X^2(t, r) dr^2 + r^2 d\Omega^2
\end{equation}

where $\alpha(t, r)$ is the lapse function, $X(t, r) = \left(1 - \frac{2GM(t,r)}{r}\right)^{-1/2}$ and $M(t,r)$ is the enclosed gravitational mass. In the specific case, we include the radial gauge, polar slicing metric $g_{ab} = diag(-\alpha^2, X^2, R^2, R^2\, sin\theta^2)$. Details appear in appendix \ref{app:rgps}. In this setting, effectively, deconfinement is triggered when the NS central density \( \rho_c(t) \) exceeds a critical threshold \( \rho_{\mathrm{crit}} \), consistent with values obtained in hybrid EOS constructions and perturbative QCD estimates \citep{Sagert+09,Ivanytskyi_2019,FischerT+11}. 

The central density evolution in the leakage-based evolution is accurately captured by the profile
\begin{equation}
    \rho_c(t) = \rho_{\mathrm{base}} + (\rho_{\mathrm{max}} - \rho_{\mathrm{base}}) \cdot \frac{1}{2} \left[1 + \tanh\left( \frac{t - t_0}{\delta t} \right) \right],
\end{equation}
where $\rho_{\mathrm{base}}$, $\rho_{\mathrm{max}}$ are base and maximum densities reached in $\delta t = 0.01 \, \mathrm{s}$. The central time $t_0 = 0.02 \, \mathrm{s}$ reflects the approximate onset of rapid collapse. Solving numerically the condition $\rho_c(t) = \rho_{\mathrm{crit}}$, one finds that the quark matter core begins forming at $t_{\mathrm{quark}} \approx 0.045$ s, which sets the center of the transition in our volume-growth function. This timing agrees with more advanced core-collapse simulations finding similar timescales after bounce, particularly in massive progenitors or EOSs with stiff phases 
\citep{Nakazato+08,DasguptaB+10}.

\begin{figure*}[t]
    \centering
    \includegraphics[width=0.99\linewidth]{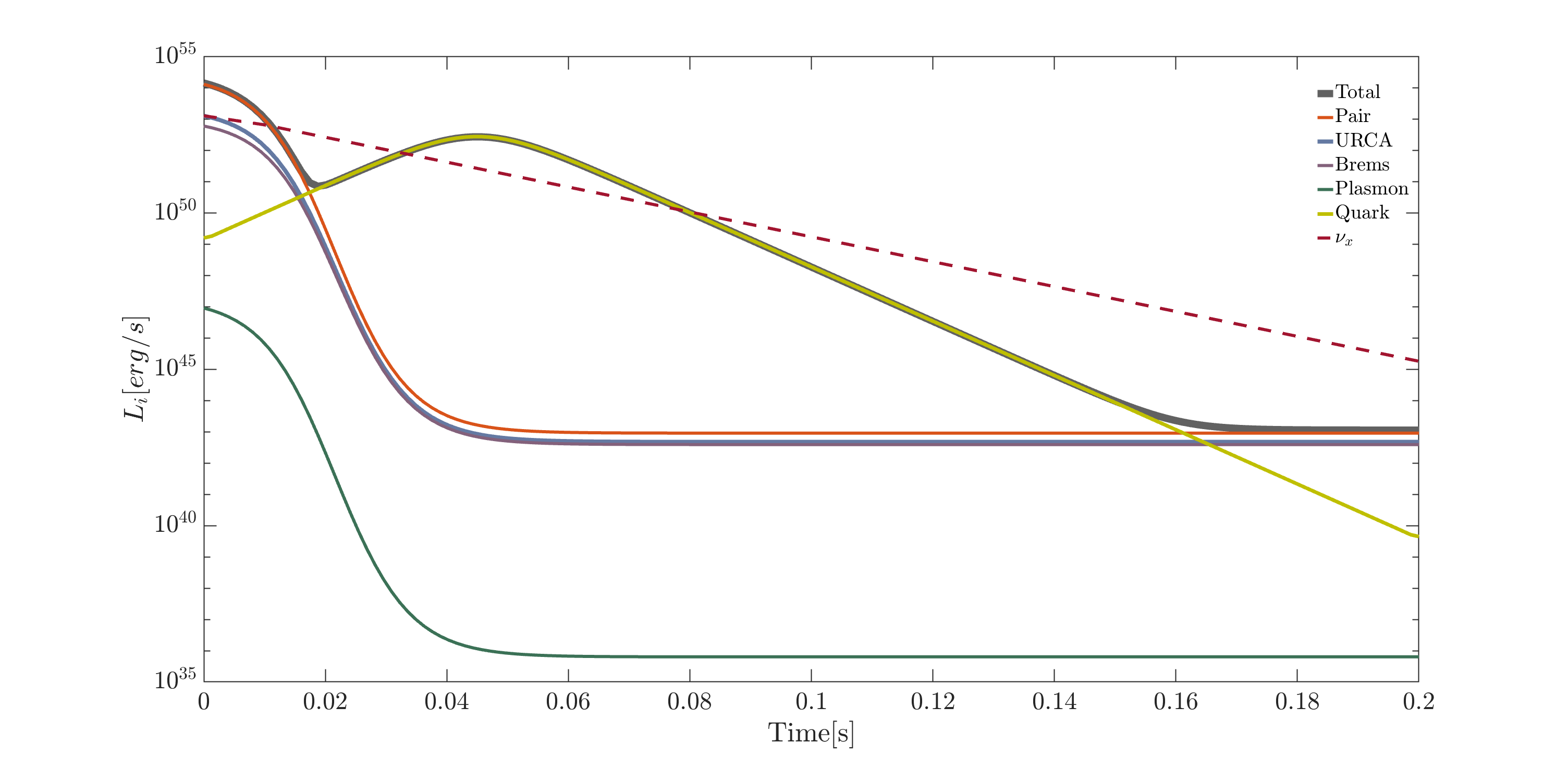}

    \caption{Time evolution of individual neutrino luminosity components $L_i$ during the collapse of a NS to a QS from channels in Table \ref{tab:table1}. Exponential suppression due to early neutrino trapping is included. The olive line represents the deconfinement burst arising from hadronic to quark matter. The purple dash line is the $\nu_x$ emission, building more gradually to peak slightly later, are due to thermal pair processes dominating.}
    \label{fig:luminosities_individual}
\end{figure*}

We compute flavor-resolved $\nu$ emission via opacity–leakage with gravitational redshift, including reactions in Table \ref{tab:table1}. Details are given in  appendix \ref{app:nu_leakage} The hadron-quark transition releases  heat on a growth timescale  generically imprinting a short $\rm \sim 10–50$ ms, spectrally harder feature in neutrino luminosities $L_\nu(t)$. In Fig.\eqref{fig:luminosities_individual} we show the individual contributions of different processes included in Table \ref{tab:table1} along with exponential suppression due to early neutrino trapping phase. Olive line represents the deconfinement burst arising from hadronic to quark matter. All of them add up in the thicker line labeled as Total. We also plot for comparison the $\nu_x$ (heavy-flavor neutrino), as we obtain from our simulation. %Heavy-lepton neutrinos result in a delayed peak with respect to the electron flavors.

%%%%%%%%%%%%%%%%%%%%%%%%%%%%%%%%%%%%%%%%%%
%%%%%%%%%%%%%%%%%%%%%%%%%%%%%%%%%%%%%%%%%

From inspection, three distinctive features emerge. First is the presence of an elevated peak/plateau ratio $R_{\rm pp}$, second, a lag $\Delta t$ among the double burst in luminosity decline-rise-decline correlated with the central-density rise. %and, last, a flavor hardening $\Delta\!\langle E_\nu\rangle$. 
The double burst has been obtained in Supernova simulations undergoing such QCD transitions, see \cite{FischerT+11, lin2023detectabilityneutrinosignalfluctuationsinduced} or in previous works on NS with seeds of quark matter \cite{Herzog_Ropke11,Pagliara+13,Drago_2016}.

%\apgcomm{We must discuss flavor ratios, as this is important. Other earlier works already gave Luminosities.} \yossefcom{Check below }\catcom{Note that nux and nue should be delayed as nux escape more easily, leaving the star earlier.}

Flavor emission in neutrino dynamics is ultimately determined by composition. Bulk emission is taken into account with the effective emissivities, as discussed. However, at lower densities, flavor oscillation becomes possible from the Mikheyev–Smirnov–Wolfenstein (MSW) effect \cite{WolfensteinL78, MikheyevP_SmirnovYu85,SmirnovYu2005_MSW, FarzanY_TortolaM18}. 

 In our simulation this is accounted for assuming adiabatic conditions as neutrinos leave the star so that we apply a simplified procedure with constant survival probabilities for electron neutrinos
 
\begin{equation}\label{survival}
    P_{\nu_e}^{\mathrm{NH}} =\left|U_{e 2}\right|^2 \approx 0.3; \quad P_{\nu_e}^{\mathrm{IH}} =\left|U_{e 1}\right|^2 \approx 0.7
\end{equation}

for the normal hierarchy (NH) and inverted hierarchy (IH) neutrino mass hierarchies, respectively, arising from standard approximate values derived from global neutrino oscillation fits (i.e. from the PMNS matrix with $\theta_{12} \approx 33^{\circ}, \theta_{13} \approx 9^{\circ}$ ). In standard three-flavour oscillations, the rest is shared mostly between $\nu_\mu$ and $\nu_\tau$. Because $\theta_{23}$ is close to $45^{\circ}$, both  equally mixed so $P\left(\nu_e \rightarrow \nu_\mu\right) \approx P\left(\nu_e \rightarrow \nu_\tau\right) \approx \frac{1-P_{\nu_e}}{2}$.

After MSW conversion, a Galactic event at $\sim$10 Kpc is resolvable in ice- and water-based Cherenkov detectors such as IceCube/Hyper-K, mainly sensitive to $\bar{\nu}_e$ via inverse beta decay on free protons. We discuss in what follows that neutrino signal from a NS to QS transition event should be distinguishable from NS formation with no-HQPT collapse. This emission would come out jointly with a kHz GW $f_2$ outlier, these observables provide a concrete multimessenger pathway to identify deconfinement at supranuclear density.

We provide order-of-magnitude estimates for the number of events observable in a Cherenkov detector such as IceCube at $d=10\ \rm Kpc$, given by
\begin{equation}
N_{\rm ev}
\simeq
\frac{E_{\nu,\rm tot}}
{4\pi d^2\langle E_\nu\rangle}
\,
\langle \sigma_\nu \rangle
\,
\epsilon_{\rm eff}
\,
N_{\rm target}
\end{equation}
with $\epsilon_{\rm eff}\sim 0.8$--$0.9$ for Super-K and Hyper-K, assuming $\langle E_\nu \rangle = 15\,\mathrm{MeV}$, $\sigma_\nu \sim 10^{-41}\,\mathrm{cm}^2$, and $N_{\mathrm{target}} \sim 10^{38}$; from our simulations including MSW effects we obtain $\langle E^{\rm NH}_\nu \rangle \sim 15\,\mathrm{MeV}$ and $\langle E^{\rm IH}_\nu \rangle \sim 12\,\mathrm{MeV}$. 
From Fig.\eqref{fig:luminosities_individual} and considering that the event rate scales as $1/d^2$, implying that a HQPT in Andromeda ($d\simeq 770\,\mathrm{Kpc}$) would yield $\sim 6000$ times fewer events than at $10\,\mathrm{kpc}$. 
Thus, a signal corresponding to $\sim 10^5$ events in Hyper-K for a Galactic source would be reduced to only $\mathcal{O}(10)$ events at the distance of Andromeda.
%The blue curve is illustrate regular dens matter of NS collapse without quark transition.

%%%%%%%%%%%%%%%%%%%%%%%%%%%%%%%%%%%%
\begin{figure*}[t]
    \centering
     \includegraphics[width=0.99\textwidth]{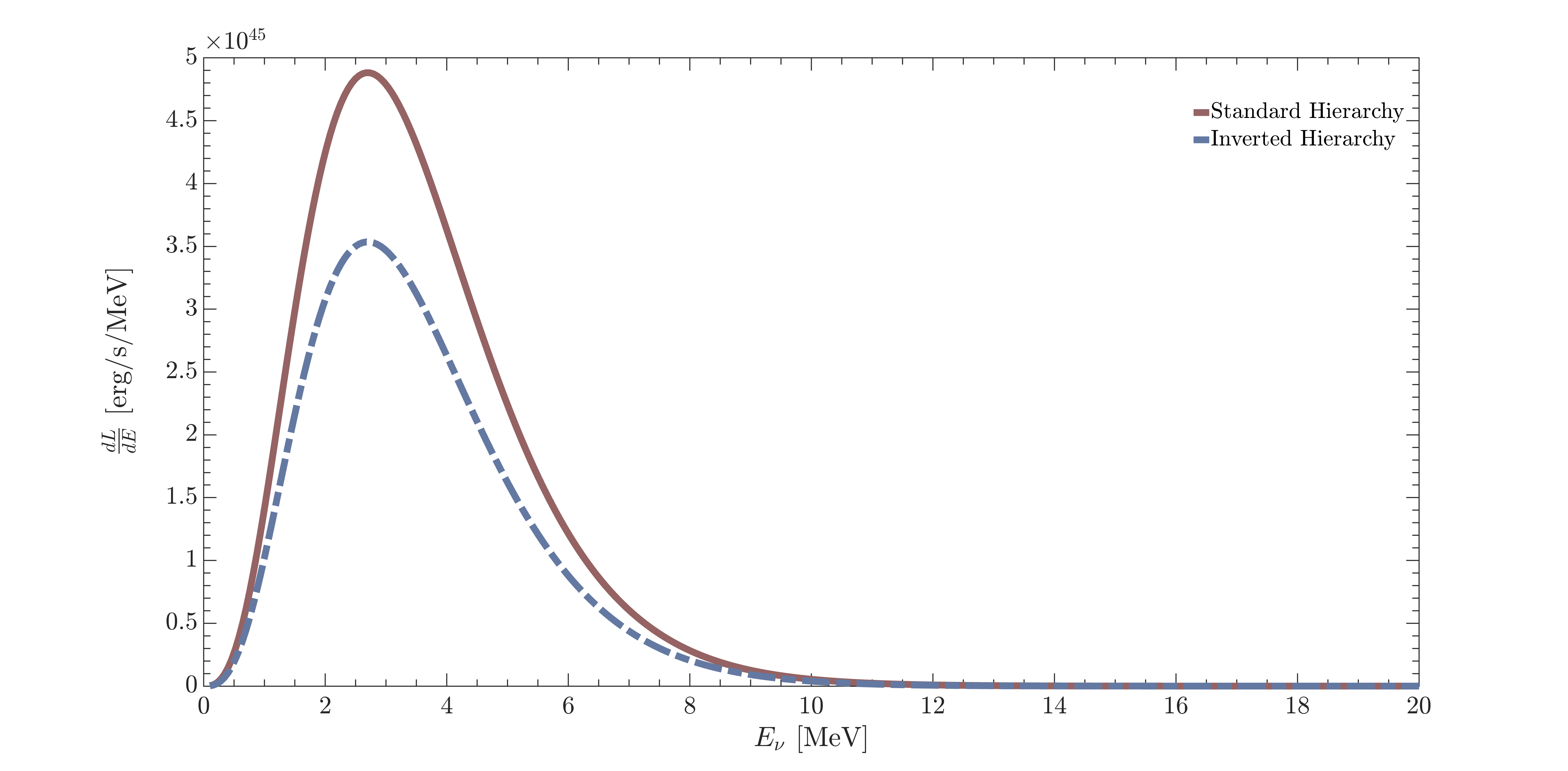}
    \includegraphics[width=0.99\textwidth]{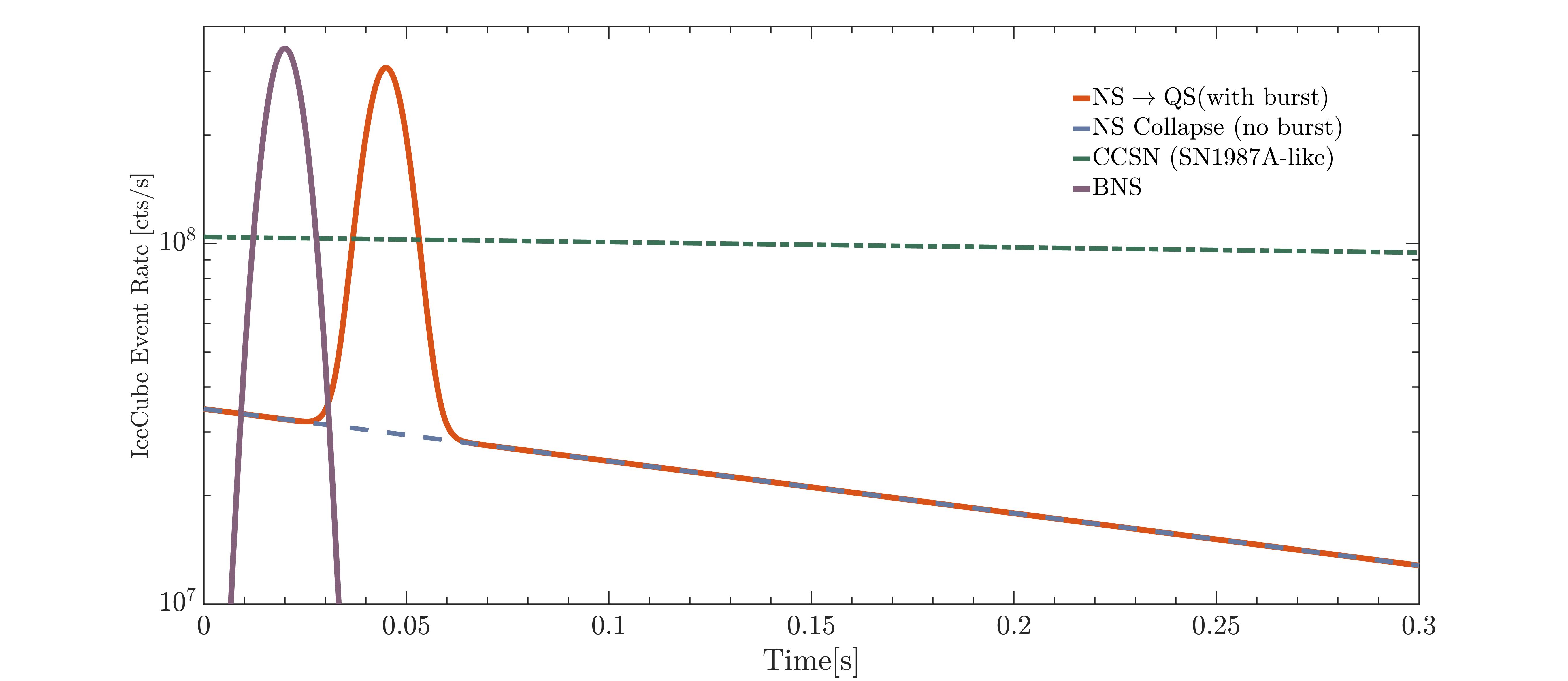}
    \caption{ Upper panel: Electron \emph{anti}-neutrino differential luminosity after MSW flavor conversion for the normal hierarchy (NH, solid red) and inverted hierarchy (IH, blue dashed). Bottom panel: Predicted IceCube detection rates from various  scenarios at a distance of $d = 10$ Kpc. The solid orange curve shows HQPT at ($t \approx 45\,\mathrm{ms}$), which induces a sharp neutrino burst, in deep contrast to the dashed blue line in absence of it. Green dash line represents a canonical CCSN; SN1987A-like, while the purple curve represents a short, high-luminosity signal from a $2.8M_\odot$ BNS.}
    \label{fig:Icecube_sig}
\end{figure*}
%%%%%%%%%%%%%%%%%%%%%%%%%%%%%%%%%%%%%%%
In Fig. \eqref{fig:Icecube_sig} we show (upper panel) the electron \emph{anti}-neutrino differential luminosity after MSW flavor conversion under the adiabatic approximation, shown for the NH (solid red) and IH (blue dashed), The post-MSW spectrum is constructed as $\mathrm{d}L_{\bar\nu_e}/\mathrm{d}E = P_{\nu_e}\,(\mathrm{d}L_{\bar\nu_e}/\mathrm{d}E)_{\rm src} + (1-P_{\nu_e})\,(\mathrm{d}L_{\nu_x}/\mathrm{d}E)_{\rm src}$ with survival probabilities $P_{\nu_e}$ as in Eq.\eqref{survival}.

In the bottom panel we show, for comparison, the predicted IceCube neutrino detection rates from various compact object collapse scenarios at an assumed distance $d = 10$ Kpc. The possibly detectable signal in IceCube must be understood as a collective excess of photomultiplier rates at the level of $\sim 10^7$--$10^8\,\mathrm{s^{-1}}$, rather than individually reconstructed neutrino events. The solid orange curve shows a NS collapse to a QS, including a quark phase transition at ($t \approx 45\,\mathrm{ms}$), which induces a sharp neutrino burst. In detail, this refers to the collapse of a ($1.8M_\odot$) NS modeled with the DD2 EOS at fixed temperature $T = 10\ \rm MeV$ and electron fraction ($Y_e = 0.2$). The orange curve shows the contribution from the quark burst alone. Neutrino trapping is included via an exponential optical depth suppression, as discussed earlier in the manuscript. As observed, the effect of HQPT leads to a sharp, transient increase in neutrino luminosity, which may serve as an observable signature of QCD phase transitions in the collapse.

This is in deep contrast to the dashed blue in absence of such HQPT. For comparison, the green dash line represents a canonical CCSN, SN1987A-like, while the purple curve represents a short, high luminosity signal from a BNS with a total mass around $\rm \sim2.8M_\odot$, as a normalization reference value. The peak of the red curve marks the onset of the quark transition. This impulsive peak offers a potentially unique observational signature of quark deconfinement in dense nucleon matter.

{\bf Conclusions.--} We have investigated the diagnostics in the neutrino emission associated with a hadron--quark phase transition (HQPT) occurring in a warm neutron star that converts into a quark star. Within a meaningful microphysics-based framework, we model the dominant neutrino production channels in dense matter, including nucleonic emission processes together with quark reactions  that become operative once deconfinement takes place. Leptonic sectors are added to the scenario as well.

Our calculations indicate that the onset of deconfinement can imprint characteristic time-dependent features in the neutrino luminosity. In particular, the transition may produce a short-lived enhancement in the neutrino flux on millisecond timescales, appearing either as a secondary peak or as a shoulder in the luminosity curve following the initial rise of the signal. This behavior reflects the rapid thermodynamic rearrangement of the stellar core during the conversion and the change in the underlying weak interaction channels once quark matter becomes energetically favored.

When compared with typical neutrino signals expected from CCSN and BNS, the predicted signal from NS to QS conversion events shows a distinct temporal structure. For a Galactic event, the corresponding neutrino luminosities may fall within the sensitivity range of current large-volume detectors such as IceCube and could be further probed by next-generation observatories including Hyper-Kamiokande. In this context, the identification of delayed maxima or transient spectral hardening in the neutrino signal could provide a useful diagnostic of the conversion process, particularly when combining detectors sensitive to different neutrino flavors and interaction channels.

The purpose of the present work is not to provide a fully consistent neutrino--radiation hydrodynamical simulation of the conversion dynamics, but rather to identify qualitative diagnostics that may arise in the neutrino signal when deconfinement occurs. Future studies incorporating alternative equations of state, improved neutrino transport, and multidimensional effects will be required to assess the robustness of these signatures.

\begin{acknowledgments}
YZ acknowledge support from MAOF grant 12641898. CA and MAPG acknowledge  partial finantial support by Junta de Castilla y León SA101P24, SA091P24,  MICIU project PID2022-137887NB-I00, Gravitational Wave Network (REDONGRA) Strategic Network (RED2024-153735-E) from Agencia Estatal de Investigación del Ministerio de Ciencia, Innovación y Universidades (MICIU/AEI/10.13039/501100011033) and COST Action COSMIC WISPers CA21106.
\end{acknowledgments}

%\appendix

\section*{Appendix}
\label{sec:appA}

\subsection{Spacetime, kinematics, and hydro variables (1D RGPS gauge)}
\label{app:rgps}
We adopt radial-gauge polar-slicing (RGPS) in spherical symmetry with vanishing shift,
\begin{equation}
  ds^2 = -\alpha^2(t,r)\,dt^2 + X^2(t,r)\,dr^2 + r^2 d\Omega^2 ,
\end{equation}
where $\alpha$ is the lapse and $X$ is the radial metric factor. The Eulerian (normal) observer is
\begin{equation}
  n^\mu=(1/\alpha,0,0,0),\qquad n_\mu=(-\alpha,0,0,0).
\end{equation}
The spatial metric is $\gamma_{ij}=\mathrm{diag}(X^2,r^2,r^2\sin^2\theta)$.
\\
\paragraph{Velocity conventions.}
In 3+1 form the fluid 4-velocity is written as $u^\mu=W(n^\mu+v^\mu)$ with $v^\mu n_\mu=0$.
For purely radial motion ($v^\theta=v^\phi=0$) we use the Eulerian radial 3-velocity $v^r$, and define
\begin{equation}
  v^2 \equiv \gamma_{rr}(v^r)^2 = X^2(v^r)^2,\,
  v \equiv X\,v^r,\,
  W=(1-v^2)^{-1/2}.
\end{equation}
The covariant radial component is $v_r\equiv \gamma_{rr}v^r=X^2 v^r=Xv$. The 4-velocity components are
\begin{equation}
  u^\mu=\left(\frac{W}{\alpha},\,W v^r,\,0,\,0\right)
      =\left(\frac{W}{\alpha},\,\frac{W v}{X},\,0,\,0\right).
\end{equation}\\

\paragraph{Perfect-fluid stress--energy and enthalpy}
The matter stress--energy tensor is

\begin{equation}
  T^{\mu\nu}_{\rm m} = (\rho h)\,u^\mu u^\nu + P\,g^{\mu\nu},
\end{equation}
where $\rho$ is the baryon rest-mass density, $P$ the pressure, and
\begin{equation}
  h = 1+\epsilon_{\rm int}+\frac{P}{\rho}
\end{equation}
is the specific enthalpy (geometric units $G=c=1$).\\

\paragraph{Conservative variables.}
We adopt the standard 1D Valencia-like conserved variables (adapted to RGPS),

\begin{equation}
  D   = \rho W,\qquad
  S_r = \rho h W^2 v_r,\qquad
  \tau= \rho h W^2 - P - D,
\end{equation}
and the electron-fraction density

\begin{equation}
  D Y_e \equiv D\,Y_e .
\end{equation}

\paragraph{Evolution system}
In this work the evolution includes approximate neutrino-matter coupling through
(i) an energy-loss term (cooling/heating) in the energy equation and (ii) a charged-current
lepton-number source in the $Y_e$ equation, computed from the leakage microphysics described
in Appendix~\ref{app:nu_leakage}. Momentum exchange is neglected in the present leakage
approximation (isotropic emission in the comoving frame.%; see Appendix~\ref{app:nu_leakage}).

Schematically, the conservative evolution has the form
\begin{align}
  \partial_t \mathbf{U} + \frac{1}{r^2}\partial_r\!\left(\frac{\alpha r^2}{X}\,\mathbf{F}\right)
  = \mathbf{S}_{\rm geom} + \mathbf{S}_\nu ,
\end{align}
with $\mathbf{U}=(D,S_r,\tau,DY_e)$. The neutrino source vector is taken as
\begin{equation}
  \mathbf{S}_\nu = \bigl(0,\,0,\,-\alpha X\,Q_{\rm leak},\,\alpha X\,D\,\Gamma_{Y_e}\bigr),
\end{equation}
where $Q_{\rm leak}(r,t)$ is the net (escaping) neutrino energy-loss rate per unit proper volume
(positive for cooling), and $\Gamma_{Y_e}(r,t)$ is the net electron-fraction change rate per unit
proper time. Both are obtained from the leakage prescription and the local thermodynamic state
$(\rho,T,Y_e)$.\\

\paragraph{Electron-fraction source from charged-current leakage.}
Only charged-current reactions change $Y_e$. In a multigroup leakage implementation it is convenient
to write the source in terms of \emph{number} emission/absorption rates. Let
$\mathcal{R}_{\nu_e,\epsilon}$ and $\mathcal{R}_{\bar\nu_e,\epsilon}$ denote the net \emph{escaping}
number emission rates per unit proper volume (including attenuation) in each comoving energy bin
centered at $\epsilon$. Then
\begin{equation}
  \Gamma_{Y_e} = -\frac{m_N}{\rho}\sum_{\epsilon}
  \left(\mathcal{R}_{\nu_e,\epsilon}-\mathcal{R}_{\bar\nu_e,\epsilon}\right),
  \label{eq:Ye_source_leakage}
\end{equation}
where $m_N$ is the nucleon mass and $n_b=\rho/m_N$ is the baryon number density. The sign convention
in Eq.~\eqref{eq:Ye_source_leakage} corresponds to $\nu_e$ emission reducing $Y_e$ (via $e^-+p\to n+\nu_e$)
and $\bar\nu_e$ emission increasing $Y_e$ (via $e^++n\to p+\bar\nu_e$). In practice, $\mathcal{R}$ is
computed consistently with the charged-current emissivity and the escape probability defined in
Appendix~\ref{app:nu_leakage}. (If muons are not included in the composition, $Y_\mu$ is not evolved
in the present setup.)

%============================================================
\subsection{Neutrino sector: formal stress--energy vs.\ the leakage approximation}
\label{app:nu_leakage}

We now make explicit \emph{what is solved} and \emph{what is approximated}
in the neutrino treatment, to avoid confusion with full neutrino radiation hydrodynamics.

\paragraph{Formal coupled system (for completeness).}
In a fully coupled neutrino radiation-hydrodynamics system the total stress--energy is
\begin{equation}
  T^{\mu\nu}_{\rm tot}=T^{\mu\nu}_{\rm m}+\sum_s T^{\mu\nu}_{\nu,s},
\end{equation}
and matter and neutrinos exchange energy--momentum through a four-force density $G^\nu_s$:
\begin{equation}
  \nabla_\mu T^{\mu\nu}_{\rm m} = -\sum_s G^\nu_s,\qquad
  \nabla_\mu T^{\mu\nu}_{\nu,s} =  \sum_s G^\nu_s,\,\,
  \Rightarrow\ \nabla_\mu T^{\mu\nu}_{\rm tot}=0,
\end{equation}
with neutrino species $s\in\{\nu_e,\bar\nu_e,\nu_x\}$ (with $\nu_x$ denoting heavy-lepton flavors).\\

\paragraph{What we actually do in this work (energy-dependent leakage).}
We do \emph{not} evolve neutrino moments (no two-moment scheme with a closure) and do \emph{not} solve a
Boltzmann equation. Hence there is no evolved neutrino radiation field, no neutrino pressure, and no
spectral/angular transport solution. Instead, we compute local emission and attenuation using an
\emph{opacity-based leakage prescription} in energy groups.

At each $(r,t)$ we evaluate, for each species $s$ and comoving neutrino energy $\epsilon$,
\begin{equation}
  \eta_s(\epsilon),\qquad \kappa^{\rm a}_s(\epsilon),\qquad \kappa^{\rm sc}_s(\epsilon),
\end{equation}
where $\eta_s$ is the emissivity (energy emitted per unit proper volume, proper time, and energy),
and $\kappa^{\rm a}_s$, $\kappa^{\rm sc}_s$ are absorption and scattering opacities computed from the local thermodynamic state. We use the transport opacity
\begin{equation}
  \kappa^{\rm tr}_s(\epsilon) \equiv \kappa^{\rm a}_s(\epsilon)+\kappa^{\rm sc}_s(\epsilon),
\end{equation}
and define the optical depth (proper distance element $dl=X\,dr$)
\begin{equation}
  \tau_s(\epsilon;r,t)=\int_r^\infty \kappa^{\rm tr}_s(\epsilon;r',t)\,X(r',t)\,dr' .
  \label{eq:tau_def}
\end{equation}\\

%%%%%%%%%%%%%%%%%%%%%%%%%%%%%%%%%%%%%%%%%%%%%%%%%%%
\begin{table*}[t]
\centering
\caption{Neutrino microphysics used in the leakage model. We list contributions to emissivity $\eta$,
absorption opacity $\kappa^{\rm a}$, and scattering opacity $\kappa^{\rm sc}$ by phase and flavor. See \citep{FrimanB_MaxwellV79,Salpeter_Shapiro81, IwamotoN82_quark,MadsenJ92,MeyerS94,HannestadS_RaffeltG98,ThompsonT+00,KohriK_MineshigeS02,KohriK+05,SchmittA_ShterninP18,Rosero-Gil_LugonesG21}}
\label{tab:table1}
\begin{tabular}{lllllll}
\toprule
Phase & Process & Reaction & Flavors & $\eta$ & $\kappa^{\rm a}$ & $\kappa^{\rm sc}$ \\
\hline
Hadronic & Direct Urca &
$n\to p+e^-+\bar\nu_e$ (and inverse) &
$\nu_e,\bar\nu_e$ & yes & yes & -- \\
Hadronic & Modified Urca &
$n+n\to n+p+e^-+\bar\nu_e$ (and inverse) &
$\nu_e,\bar\nu_e$ & yes & yes & -- \\
Hadronic & $e^-e^+$ annihilation &
$e^-+e^+\to \nu+\bar\nu$ &
all (dominant $\nu_x$) & yes & -- & -- \\
Hadronic & Plasmon decay &
$\gamma^\ast\to \nu+\bar\nu$ &
all (often $\nu_x$-dominated) & yes & -- & -- \\
Hadronic & NN bremsstrahlung &
$N+N\to N+N+\nu+\bar\nu$ &
all (often $\nu_x$-dominated) & yes & -- & -- \\
Hadronic & $\nu$--$N$ scattering &
$\nu+N\to \nu+N$ &
all & -- & -- & yes \\
Hadronic & $\nu$--$e^\pm$ scattering &
$\nu+e^\pm\to \nu+e^\pm$ &
all & -- & -- & \emph{state yes/no} \\
Quark & Quark direct Urca &
$d\to u+e^-+\bar\nu_e$ (and $s$ analogs, if present) &
$\nu_e,\bar\nu_e$ & yes & yes & -- \\
Quark & $\nu$--$q$ scattering &
$\nu+q\to \nu+q$ &
all & -- & -- & yes \\
\hline
\end{tabular}
\end{table*}
%%%%%%%%%%%%%%%%%%%%%%%%%%%%%%%%%%%%%%%%%%%%%%%%%%%%%%%%%%%%
\paragraph{Escape probability}

The escaping emissivity is modeled by an escape probability $\mathcal{P}_{\rm esc}$ that reduces
emission in optically thick regions,
\begin{equation}
  \eta^{\rm esc}_s(\epsilon;r,t)=\eta_s(\epsilon;r,t)\,
  \mathcal{P}_{\rm esc}\!\left[\tau_s(\epsilon;r,t)\right],
  \, \text{e.g.}\ \mathcal{P}_{\rm esc}=e^{-\tau_s}.
  \label{eq:eta_esc}
\end{equation}

Notably, Alternative interpolations between free streaming and diffusion may be used; the specific choice
adopted in the main text should be stated consistently with Eq.~\eqref{eq:eta_esc}. The escaping energy loss rate is then

\begin{equation}
  Q_{\rm leak}(r,t)=\sum_s\int_0^\infty \eta^{\rm esc}_s(\epsilon;r,t)\,d\epsilon,
  \label{eq:Qleak_def}
\end{equation}

and the escaping \emph{number} emission rates entering Eq.~\eqref{eq:Ye_source_leakage} are
\begin{equation}
  \mathcal{R}_{s}(r,t)=\int_0^\infty \frac{\eta^{\rm esc}_s(\epsilon;r,t)}{\epsilon}\,d\epsilon,
  \qquad
  \mathcal{R}_{s,\epsilon}\approx \frac{\eta^{\rm esc}_s(\epsilon)}{\epsilon}\,\Delta\epsilon .
\end{equation}

\paragraph{Coupling to the GRHD evolution}
In the present leakage approximation the neutrino four-force is taken to be purely timelike in the
fluid rest frame (isotropic emission), so that momentum exchange is neglected:
\begin{equation}
  G^\mu \approx Q_{\rm leak}\,u^\mu ,
\end{equation}
which yields the energy sink term $-\alpha X Q_{\rm leak}$ in the conservative energy equation used
in Appendix~\ref{app:rgps}. Lepton-number exchange is included by evolving $DY_e$ with the charged-current
source $\Gamma_{Y_e}$ computed from the leakage number rates in Eq.~\eqref{eq:Ye_source_leakage}.\\

\paragraph{Luminosity at infinity (redshift + proper volume).}
The observable luminosity at infinity is obtained by integrating the escaping emission over proper volume
and applying gravitational redshift,
\begin{equation}
  L_{s,\infty}(t)=\int_0^\infty 4\pi r^2 X(r,t)\,\alpha^2(r,t)\,Q^{\rm esc}_s(r,t)\,dr,
  \label{eq:L_infty}
\end{equation}
where $Q^{\rm esc}_s$ is the escaping energy emission rate per unit proper volume for species $s$
(obtained by integrating $\eta^{\rm esc}_s$ over $\epsilon$), and $dV_{\rm proper}=4\pi r^2 X\,dr$.
The factor $\alpha^2$ accounts for the redshift of energies and rates between the emitting fluid frame
and an observer at infinity.\\

\paragraph{Scope and limitations}
Because we do not solve for neutrino moments or a Boltzmann equation, the present treatment does not 
include multi-angle transport, self-consistent spectral formation by scattering/thermalization, neutrino
pressure/viscosity contributions to the dynamics, or momentum exchange with the fluid. Any omissions in the
opacity set (e.g. neutrino--$e^\pm$ scattering, if not implemented) should be stated explicitly, since they
can bias spectral tails and luminosities. The purpose of this leakage-based model is therefore to provide a
transparent, microphysics-forward estimate of time-dependent neutrino emission associated with the HQPT,
rather than a state-of-the-art neutrino radiation-hydrodynamics prediction.

\subsection{Microphysics summary}
\label{app:microphysics_table}

For reproducibility, we summarize in Table \ref{tab:table1} the neutrino processes included in the leakage source terms.
The emissivities/opacities are evaluated separately in hadronic matter and in quark matter (MIT bag),
using the local thermodynamic state and composition.

\subsection{Warm quark MIT bag model}
\label{app:mitbag}

For the MIT bag model, quark chemical potentials are written in terms of conserved-charge $\mu_B,\mu_Q$ baryonic and quark potentials, respectively 

\begin{equation}
  \mu_u=\frac{1}{3}\mu_B+\frac{2}{3}\mu_Q,\qquad
  \mu_d=\frac{1}{3}\mu_B-\frac{1}{3}\mu_Q,\qquad
  \mu_s=\frac{1}{3}\mu_B-\frac{1}{3}\mu_Q.
\end{equation}
Quark number densities $n_f$, ${f=u,d,s}$, pressures $P_f$, and energy densities $\varepsilon_f$ follow from
finite-$T$ Fermi integrals $n_F(E ; \mu, T)=\frac{1}{e^{(E-\mu) / T}+1}$

\begin{align}
n_f(\mu_f,T) &=
\frac{g_f}{2\pi^2}\int_0^\infty p^2
\left[
n_F(E;\mu_f,T)-n_F(E;-\mu_f,T)
\right] dp,\\[4pt]
P_f(\mu_f,T) &=
\frac{g_f}{6\pi^2}\int_0^\infty \frac{p^4}{E}
\left[
n_F(E;\mu_f,T)+n_F(E;-\mu_f,T)
\right] dp,\\[4pt]
\varepsilon_f(\mu_f,T) &=
\frac{g_f}{2\pi^2}\int_0^\infty p^2 E
\left[
n_F(E;\mu_f,T)+n_F(E;-\mu_f,T)
\right] dp .
\end{align}

with $E=\sqrt{p^2+m_f^2}$ and degeneracy $g_f=6$. The total quark pressure and energy density are
\begin{equation}
  P_Q = \sum_{f=u,d,s}P_f - B,\qquad
  \varepsilon_Q = \sum_{f=u,d,s}\varepsilon_f + B,
\end{equation}
and the conserved-charge densities used by the Gibbs solver are
\begin{equation}
  n_{B,Q}=\frac{n_u+n_d+n_s}{3},\qquad
  n_{Q,Q}=\frac{2}{3}n_u-\frac{1}{3}n_d-\frac{1}{3}n_s.
\end{equation}

%\subsection{Metric and Coordinate Choice} \label{subsec:Metric}

%\nocite{*}

\bibliography{apssamp}% Produces the bibliography via BibTeX.
%\bibliography{sorsamp}
\end{document}